\documentclass[a4paper]{jpconf}
\usepackage{graphicx, epsfig, psfig}

\def\spose#1{\hbox to 0pt{#1\hss}}
\def\ltapprox{\mathrel{\spose{\lower 3pt\hbox{$\mathchar"218$}}
 \raise 2.0pt\hbox{$\mathchar"13C$}}}
\def\gtapprox{\mathrel{\spose{\lower 3pt\hbox{$\mathchar"218$}}
 \raise 2.0pt\hbox{$\mathchar"13E$}}}

\begin{document}

\title{Simulating a CP-violating topological term in gauge theories}

\author{Haralambos Panagopoulos\,$^{a*}$ and Ettore Vicari\,$^b$}

\address{$^a$Department of Physics, University of Cyprus, Lefkosia, CY-1678, Cyprus}

\address{$^b$Dipartimento di Fisica, Universit\`a di Pisa and INFN, I-56127 Pisa, Italy}

\address{$^*$Presented the talk}

\ead{haris@ucy.ac.cy, vicari@df.unipi.it}

\begin{abstract}
We present recent results on the $\theta$-dependence of
four-dimensional SU(N) gauge theories, where $\theta$ is the
coefficient of the CP-violating topological term in the
Lagrangian. In particular, we study the scaling behavior of these theories, by Monte Carlo simulations
at imaginary $\theta$.  The
numerical results provide good evidence of scaling in the continuum
limit.  The imaginary $\theta$ dependence of the ground-state energy
turns out to be well described by the first few terms of related
expansions around $\theta=0$, providing accurate estimates of the
first few coefficients, up to $O(\theta^6)$. 
\end{abstract}

Four-dimensional $SU(N)$ gauge theories have a nontrivial dependence
on the parameter $\theta$ which appears in the Euclidean Lagrangian as
\vskip -3mm
\begin{equation}
{\cal L}_\theta  = (1/4) F_{\mu\nu}^a(x)F_{\mu\nu}^a(x)
- i\, \theta\, q(x),\qquad q(x)= g^2/(64\pi^2) \,\epsilon_{\mu\nu\rho\sigma}
F_{\mu\nu}^a(x) F_{\rho\sigma}^a(x),
\label{lagrangian}
\end{equation}
where 
$q(x)$ 
is the topological charge density.  The 
ground-state energy density $F(\theta)$ behaves as
\vskip -3mm
\begin{equation}
{\cal F}(\theta)\equiv
F(\theta)-F(0)= (1/2)\, \chi \theta^2 s(\theta),\label{ftheta}
\end{equation}
where $\chi$ is the topological susceptibility at $\theta=0$, 
\vskip -3mm
\begin{equation}
\chi \equiv {\textstyle \int} d^4 x \langle q(x)q(0) \rangle_{\theta=0} 
= \langle Q^2 \rangle_{\theta=0} / V,\quad
Q\equiv {\textstyle \int} d^4x\, q(x),
\label{chidef}
\end{equation}
$V$ is the spacetime volume and $s(\theta)$ is a dimensionless even
function of $\theta$ such that 
$s(0)=1$.  Assuming analyticity at $\theta=0$, $s(\theta)$ can be
expanded as: $s(\theta) = 1 + b_2 \theta^2 + b_4 \theta^4 + \cdots,$ 
where only even powers of $\theta$ appear.  Large-$N$ scaling
arguments applied to the
Lagrangian (\ref{lagrangian}) indicate
that the relevant scaling variable in the large-$N$ limit is
$\bar\theta\equiv {\theta/N}$.  This implies that in this
limit $\chi=O(1)$, while the coefficients $b_{2i}$ are suppressed by
powers of $N$, i.e. $b_{2i}=O(N^{-2i}).$

Due to the nonperturbative nature of the $\theta$ dependence,
quantitative assessments have largely focused on the
lattice formulation of the theory, using Monte Carlo (MC) simulations.
However, the complex nature of the $\theta$ term in the Euclidean
Lagrangian prohibits a direct MC simulation at $\theta\ne 0$.
Information on the $\theta$ dependence of physically relevant
quantities, such as the ground state energy and the spectrum, has been
obtained by computing the coefficients of the corresponding expansions
around $\theta = 0$.  The coefficients of $s(\theta)$
can be determined from appropriate
zero-momentum correlation functions of $q(x)$ at $\theta=0$, which are related to the
moments of the $\theta=0$ probability distribution $P(Q)$ of the
topological charge $Q$.  Indeed
\vskip -3mm
\begin{equation}
b_2 = - {1\over 12\,\chi V} \left[ 
\langle Q^4 \rangle - 3  \langle Q^2 \rangle^2 \right]_{\theta=0}, 
\quad
b_4 = - {1\over 360\, \chi V} \left[ 
\langle Q^6 \rangle  -  15 \langle Q^2 \rangle \langle Q^4 \rangle  +
30 \langle Q^2 \rangle^3 \right]_{\theta=0},  \label{b4chi6} 
\end{equation}
etc. They
parameterize the deviations of $P(Q)$ from a simple Gaussian behavior.
It has been shown that correlations
involving multiple insertions of the topological charge
can be defined in a nonambiguous, regularization independent
way, and therefore $b_{2i}$ are well
defined renormalization group invariant quantities.  The numerical
evidence for a nontrivial $\theta$-dependence, obtained through MC
simulations, appears quite robust. 
We refer the reader to Ref.~\cite{VP-09} for a recent review.
On the other hand, MC simulations at $\theta=0$ have only made it
possible to estimate  
the ground-state energy up to $O(\theta^4)$.
The large-$N$ prediction $b_2=O(N^{-2})$ has been already supported
by numerical results; the calculation of the higher-order terms would
provide a further check of the power-law suppression predicted by
large-$N$ arguments.

In this paper we consider imaginary values of $\theta$, which make the
Euclidean Lagrangian (\ref{lagrangian}) real, thus making MC simulations possible. 
For further details and references, the reader may consult our
Ref.~\cite{PV-11}. 
Assuming analyticity at $\theta=0$, the results provide
quantitative information on the expansion around $\theta=0$.  Indeed, fits of the data
to polynomials of imaginary $\theta$ may provide more accurate
estimates of the coefficients, overcoming the rapid increase of
statistical errors observed at $\theta=0$.  Perturbative
renormalization-group (RG) arguments indicate that
$\theta$ is a RG invariant parameter of the theory, thus the continuum
limit should be approached while keeping $\theta$ fixed to any complex
value.  We find that this is indeed supported by the
numerical data for the 4D SU(3) lattice gauge theory,
at least for $|\theta|<\pi$, which are well described
by the first few nontrivial terms of the expansion around $\theta=0$.

Introducing the real parameter $\theta_i$, defined by
$\theta \equiv - i \theta_i,$ 
Eq.~(\ref{ftheta}) leads to:
\vskip -6mm
\begin{eqnarray}
&& {\langle Q \rangle_{\theta_i}\over V} = -{\partial
{\cal F}(-i\theta_i)\over \partial \theta_i} = \chi \theta_i \left( 1 - 2
b_2 \theta_i^2 + 3 b_4 \theta_i^4 + \cdots\right),
\label{qthetai}\\
&& {\langle Q^2 \rangle^{c}_{\theta_i} \over V} \equiv {\langle Q^2
\rangle_{\theta_i} - \langle Q \rangle_{\theta_i}^2 \over V} =
-{\partial^2 {\cal F}(-i\theta_i)\over \partial \theta_i^2} = \chi \left( 1
- 6 b_2 \theta_i^2 + 15 b_4 \theta_i^4 + \cdots\right).
\label{chithetai}
\end{eqnarray}
\vskip -3mm

\begin{figure}[t]
\hspace{4pc}\includegraphics[width=14pc]{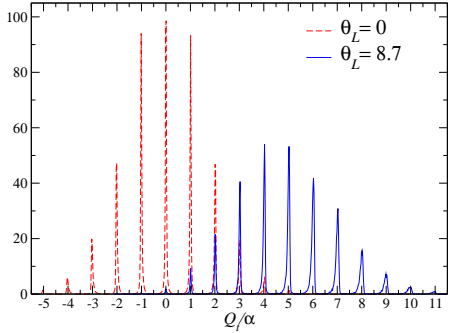}\hspace{2pc}%
\begin{minipage}[b]{14pc}\caption{\label{histogram}Distribution of the
    ratio $Q_t/\alpha$, for $\beta=6.2$ 
configurations at $\theta_L = 0$ and $\theta_L=8.7$
($\theta_i\approx 1.5$).\\
\phantom{a}}
\end{minipage}
\vskip -0.5cm
\end{figure}

The nonperturbative formulation of the above theory on the lattice
requires a discretization of the action, $S_L - \theta_L Q_L$; for
$S_L$ we use the plaquette gluon action, while for $Q_L$
we employ the ``twisted double plaquette'' operator $q_L$ ($Q_L =
\sum_x q_L(x)$).
Notice that this is not the only possible choice for $q_L$\,; the only
requirement 
is that it have the correct continuum limit when
$a\to 0$ ($a$: lattice spacing). In the continuum limit $q_L(x)$, being
a local operator, behaves as
\vskip -3mm
\begin{equation}
q_{L}(x)\longrightarrow a^4 Z_q\, q(x) + O(a^{6}),
\label{renorm}
\end{equation}
where $Z_q$ is a finite function of the bare coupling $g_0$, going to one in
the limit $\beta\equiv 2N/g_0^2 \rightarrow \infty$.  Thus, we have
the correspondence: $\theta_i = Z_q \,\theta_L,$
apart from $O(a^2)$ corrections.  The renormalization $Z_q$ may be
evaluated by MC simulation at $\theta=0$, computing
\vskip -3mm
\begin{equation}
Z_q = \langle Q Q_L \rangle_{\theta=0} / \langle Q^2 \rangle_{\theta=0} 
\,,
\label{zqc}
\end{equation}
where $Q$ is an estimator such as those obtained by the
overlap method or the cooling
method, which are not affected by
renormalizations, nor by nonphysical contact terms. Thus, the ratios
\vskip -5mm
\begin{equation}
\langle Q \rangle_{\theta_i}/ \langle Q^2 \rangle_{\theta=0}=
\theta_i  - 2 b_2 \theta_i^3 + 3 b_4 \theta_i^5 + ... ,\qquad
\langle Q^2 \rangle^c_{\theta_i}/ \langle Q^2 \rangle_{\theta=0}
= 1 - 6 b_2 \theta_i^2 + 15 b_4 \theta_i^4 + ... ,\label{avq2}
\end{equation}
are expected to have a well defined continuum limit as functions of $\theta_i$\,.

We have carried out MC simulations of the 4D SU(3)
lattice gauge theory, at $\beta=5.9,\,6,\,6.2$, for lattice sizes
$L=16,\,16,\,20$, respectively; the simulations are carried out both
at $\theta_L=0$ and $\theta_L\ne 0$, within the region
$|\theta_i|\ltapprox\pi$. Since our numerical study requires
high-statistics MC simulations, we choose the cooling method as
estimator of the topological charge $Q$. The topological charge
has been measured on cooled configurations (by locally minimizing the
lattice action), using the twisted double plaquette operator.  As is
well known, this procedure leads to 
values $Q_t \simeq k \alpha$, where $k$ is an integer and $\alpha
\ltapprox 1$.  Once we determine $\alpha$, 
we assign to $Q$ the integer closest to
$Q_t/\alpha$. This cooling method for estimating $Q$, though less
rigorous than the significantly more
expensive overlap method, produces results in good agreement with it. 

Fig.~\ref{histogram} shows the
distributions of the ratio $Q_t/\alpha$ of $\beta=6.2$ cooled
configurations at $\theta_L = 0$ and $\theta_L = 8.7$
($\theta_i=Z_q\theta_L\approx 1.5$).
We note that these distributions cluster around integer
values, also for rather large values of $Q_t/\alpha$, both for $\theta_L=0$
and $\theta_L=8.7$.

MC simulations at $\theta=0$ were performed at $\beta= 5.9,\, 6.0,\, 6.2$. 
Over 40 million sweeps per value of
$\beta$ were produced.
The results for $\chi$, $b_2$, $b_4$ and $Z_q$ are reported in Ref.~\cite{PV-11}.
Providing our improved estimate of high-order
coefficients, such as $b_4$, turned out to be very hard in $\theta=0$
MC simulations, requiring huge statistics. The results for $b_4$ are
consistent with zero, 
suggesting the bound $|b_4|\ltapprox 0.005$, which is improved in
$\theta\ne 0$ runs.
Our estimates of $Z_q$ (e.g. $Z_q(\beta = 6.0) = 0.135(1)$\,) reduce
the uncertainty on $Z_q$ as produced by other methods.

\begin{figure}[t]
\hspace{1pc}
\begin{minipage}{16pc}
\includegraphics[width=16pc]{qtheta.eps}
\vskip -5mm
\caption{\label{qtheta}$\langle Q \rangle_{\theta_L}/\langle Q^2
\rangle_{\theta=0}$ vs $\theta_i=Z_q\theta_L$. }
\end{minipage}\hspace{2pc}%
\begin{minipage}{16pc}
\includegraphics[width=16pc]{rqoq2.eps}
\vskip -5mm
\caption{\label{rqoq2}$\langle Q \rangle_{\theta_L}/ \langle Q^2
\rangle^c_{\theta_L}$ vs $\theta_i=Z_q\theta_L$.}
\end{minipage} 
\vskip -5mm
\end{figure}

MC simulations at $\theta_L\ne 0$ are slower by
approximately a factor of three,
due to the complexity of the action.
In $\theta_L\ne 0$ runs, $\sim 3$
million sweeps were produced for each value of $\beta$ and $\theta_L$.
Figs.~\ref{qtheta} and \ref{rqoq2} show results for
the ratios $\langle Q \rangle_{\theta_L}/\langle Q^2
\rangle_{\theta=0}$ and $\langle Q \rangle_{\theta_L}/\langle Q^2
\rangle^c_{\theta_L}$, versus $\theta_i=Z_q\theta_L$
(cf. Eq.~(\ref{avq2})\,).  The MC data at 
different $\beta$ values follow the same curve, providing
evidence of scaling.  Scaling corrections, expected to be
$O(a^2)$, are quite small, and tend to increase with increasing
$\theta_i$.  This good scaling behavior corroborates the
existence of a nontrivial continuum limit for any value of $\theta_i$.
Fitting our data to Eqs.~(\ref{qthetai},\ref{chithetai},\ref{avq2})
improves significantly the $\theta=0$ results.
In particular, a much smaller bound on $b_4$ is obtained: $|b_4|<
0.001$\,; also, we find: $b_2=-0.026(3)$, which is clearly more
precise than the estimate obtained from $\theta=0$ runs only:
$b_2=-0.029(7)$. 

Besides allowing more precise determinations of the $\theta$
expansion coefficients of the ground-state energy and other
observables, using imaginary $\theta$ values might turn out
useful in overcoming the dramatic critical
slowing down of topological
modes, by performing parallel
tempering simulations with a set of $\theta$ values
including $\theta{=}0$; this is an {\em exact} MC algorithm for
the model.

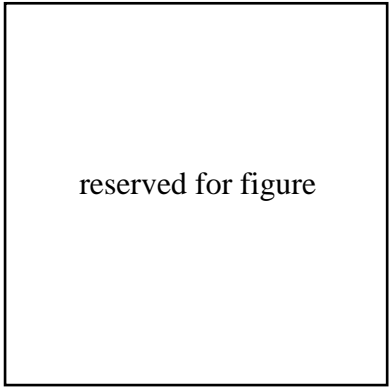
\end{document}